\documentclass[twocolumn,tighten]{aastex62}
\usepackage{amsmath}	% Advanced maths commands
\usepackage{amssymb}	% Extra maths symbols

\usepackage{color}
\usepackage[utf8]{inputenc}

\received{March 9, 2021}
\revised{---}
\accepted{April 12, 2021}
\submitjournal{ApJ}

\shorttitle{Minimum Periods of H-Rich Bodies}
\shortauthors{Rappaport et al.}

\begin{document}

\title{Minimum Orbital Periods of H-Rich Bodies}

\author[0000-0003-3182-5569]{S. Rappaport}
\affiliation{Department of Physics, and Kavli Institute for Astrophysics and Space Research, M.I.T., Cambridge, MA 02139, USA}
\correspondingauthor{S. Rappaport}
\email{sar@mit.edu}

\author[0000-0002-4870-8855]{A. Vanderburg}
\affiliation{Department of Astronomy, University of Wisconsin, Madison, WI 53715, USA}

\author[0000-0002-4870-8855]{J. Schwab}
\affiliation{Department of Astronomy and Astrophysics, University of California, Santa Cruz, CA 95064, USA}

\author[0000-0002-6916-8130]{L. Nelson}
\affiliation{Department of Physics and Astronomy, Bishop's University, 2600 College St., Sherbrooke, QC J1M 1Z7, Canada}




\begin{abstract}
In this work we derive the minimum allowed orbital periods of H-rich bodies ranging in mass from Saturn's mass to 1 $M_{\odot}$, emphasizing gas giants and brown dwarfs over the range $0.0003 - 0.074 \, M_\odot$.  Analytic fitting formulae for $P_{\rm min}$ as a function of the mass of the body and as a function of the mean density are presented. We assume that the density of the host star is sufficiently high so as not to limit the minimum period.  In many instances this implies that the host star is a white dwarf. This work is aimed, in part, toward distinguishing brown dwarfs from planets that are found transiting the host white dwarf without recourse to near infrared or radial velocity measurements.  In particular, orbital periods of $\lesssim 100$ minutes are very likely to be brown dwarfs.  The overall minimum period over this entire mass range is $\simeq 37$ minutes.
\end{abstract}


%% subject header.  Objects should be in the appropriate "individual"
%% headers (e.g. quasars: individual, stars: individual, etc.) with the
%% additional provision that the total number of headers, including each
%% individual object, not exceed six.  The \objectname{} macro, and its
%% alias \object{}, is used to mark each object.  The macro takes the object
%% name as its primary argument.  This name will appear in the paper
%% and serve as the link's anchor in the electronic edition if the name
%% is recognized by the data centers.  The macro also takes an optional
%% argument in parentheses in cases where the data center identification
%% differs from what is to be printed in the paper.



\section{Introduction}

Orbital periods in systems containing white dwarfs (`WDs') can be extremely short, especially if both of the stars are H-exhausted objects.  The best examples of this are WD+WD binaries with periods of 7 and 9 minutes \citep{burdge19,burdge20a}.  Such systems almost certainly involve one or more phases of mass transfer.  In this work, we raise the question of the minimum allowed orbital periods when at least one of the stars is still H-rich.  In this latter category are the brown dwarf (BD) plus WD binaries.  The minimum orbital periods of those systems were discussed extensively in \citet{nelson18}.  Table 1 in that paper provides a list of 25 systems with WD primaries and either a BD or lower-main-sequence (MS) companion.  These systems are thought to have had no mass transfer episodes from the BD or MS star to the WD.  The orbital periods listed there range from 250 minutes down to 68 minutes.  The minimum allowed orbital period for these WD+BD systems derived in \citet{nelson18} was $\simeq 40$ min.  

Photometric surveys like {\em Kepler}, {\em K2}, {\em TESS}, and the Zwicky Transient Facility (ZTF; \citealt{bellm17}); are discovering and will continue to discover interesting short-period binaries containing a WD (see, e.g., G.~Murawski's website\footnote{\url{https://sites.google.com/view/mgab-astronomy/eclipsing-white-dwarfs}}).  There are a number of new WD+WD binaries (see, e.g., \citealt{burdge20b}) as well as some likely WD+BD binary candidates, but none in the latter category we are aware of with periods shorter than 68 minutes.  

For a number of years now, researchers have been interested in whether planets might be found orbiting WD dwarfs (e.g., \citealt{agol11}; \citealt{lund18}; \citealt{bell19}; \citealt{cortes19}).  Such planets would have to have  survived the giant phase of the WD progenitor.  There have been a number of tantalizing suggestions in this regard.  \citet{gaensicke19} proposed that there is a disintegrating planet orbiting WDJ0914+1914 with a $\sim$ 9-d period.  More recently \citet{vanderburg20} reported the first intact transiting planet orbiting WD 1856+534 in a 1.4-day orbit. This is a gas giant planet with $M \lesssim 14 \,M_J$.

There is also evidence for dust-emitting bodies orbiting WDs.  WD 1145+017 exhibits deep dips with a characteristic period of 4.5-hour (\citealt{vanderburg15}; \citealt{gaensicke16}; \citealt{rappaport17a}).  There is also ZTF J013906.17+524536.89 with dips recurring at a $\sim$107-day period \citep{vanderbosch19}.  \citet{manser19} reported the discovery of a 123 minute periodicity in the motion of gas in the disk orbiting WD SDSS J122859.93+104032.9, which they attributed to an orbiting massive rocky body. In all these cases, the dust and gas probably originate from orbiting asteroids or planetesimals which are not the subject of this paper, and we do not consider them further.

\begin{figure}
\begin{center}
\includegraphics[width=\columnwidth]{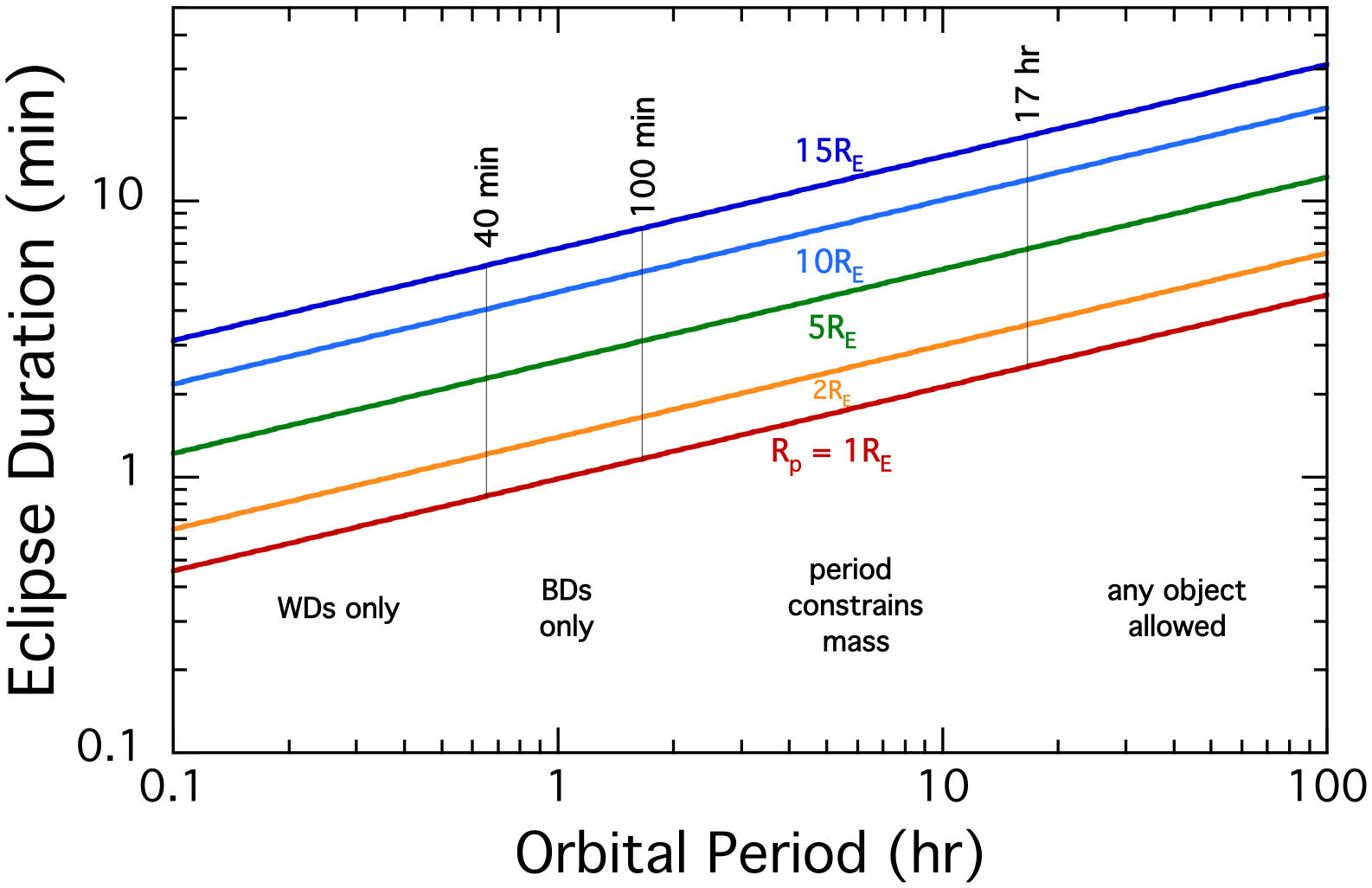}
\caption{Equatorial transit duration vs.~orbital period for objects of various sizes that are occulting a 1.4 $R_\oplus$ white dwarf.  The radii of the occulting bodies, $R_p$, are written next to each curve in units of the Earth's radius ($R_E$). Objects covering a wide range in mass from Saturn through brown dwarfs, to stars at the bottom of the main-sequence, may all have comparable radii.  In this work we explore the constraints that are set on the masses of the transiting bodies by their orbital periods.  }
\label{fig:fig1} % Figure 1
\end{center}
\end{figure}

We show in Fig.~\ref{fig:fig1} an illustrative plot of equatorial eclipse durations for objects of various masses transiting a white dwarf.  Gas giant planets to brown dwarfs, spanning 2.5 orders of magnitude in mass, and orbital periods less than a day, have a typical transit duration of $\sim$8 minutes (within a factor of a few) since all these objects can have the same radii to within $\pm 15\%$.  In this work we are interested in how knowledge of the orbital period can inform us about the nature of the transiting body.

Specifically, the goal of this paper is to identify the minimum allowed orbital period, $P_{\rm min}$, of H-rich bodies orbiting a host star as a function of their mass.  The host star is assumed sufficiently dense so as not to limit $P_{\rm min}$, i.e., to prevent the body from reaching its Roche limit (see, e.g., \citealt{roche1849}; \citealt{davidsson99}; \citealt{holsapple06}; \citealt{rappaport13}).  Orbiting bodies with masses ranging from Saturn's mass to solar-mass stars on the main sequence are considered.  In Section~\ref{sec:general} we derive analytic relations for $P_{\rm min}$ as a function of the mass and radius of the orbiting H-rich body, as well as a function of the body's mean density.  In Section~\ref{sec:rom} we give a general radius-mass relation for cold H-rich material based on Eggleton's \citep{eggleton06} analytic fitting formula.  In Section~\ref{sec:pmin} we present results for $P_{\rm min}$ as functions of the mass and of the density of the H-rich body, both in the form of graphs and analytic expressions.  We return in Sect~\ref{sec:concentration} to discuss the effect on our results of H-rich objects that are not highly centrally concentrated.  Section~\ref{sec:summary} contains a summary and our conclusions.
	
\section{Dependence of $P_{\rm min}$ on Mass and Radius}
\label{sec:general}

We start by asking two questions: (1) how close can a H-rich body be to its host star before it starts to overflow its Roche lobe, and (2) how applicable is the Roche potential to bodies like Jupiter, super-Jupiters, and brown dwarfs?  

Regarding the first question, we start by writing an expression for the size of the Roche lobe, $R_{\rm L}$, as a function of the mass ratio, $q$, of the two stars and the orbital separation, $a$, assuming a circular orbit.  This takes the form
\begin{equation}
R_{\rm L} = f(q) \, a~,
\label{eqn:rL}
\end{equation}
where we consider two well-used functions to represent $f(q)$.  Note that $4\pi R_{\rm L}^3/3$ is defined to be the volume of the Roche lobe, and we therefore we can call $R_{\rm L}$ the `volumetric radius' of the Roche lobe.  Inserting this expression into Kepler's third law, we have:
\begin{equation}
\frac{G (M_{\rm host}+M_{\rm com}) f^3(q)} {R_{\rm L}^3} = \left( \frac{2 \pi}{P} \right)^2
\label{eqn:kepler}
\end{equation}
where $M_{\rm host}$ and $M_{\rm com}$ are the masses of the host star and H-rich companion, respectively, $P$ is the orbital period, and $q \equiv M_{\rm com}/M_{\rm host}$.

An analytically convenient and reasonably accurate approximation to $R_{\rm L}$, normalized to the orbital separation, was given by \citet{kopal59} for $q < 1$:
\begin{equation}
f_{\rm K} = \frac{2}{3^{4/3}} \left( \frac{q}{1+q} \right)^{1/3},
\label{eqn:kopal}
\end{equation}
where the numerical value of the leading factor is 0.4622.  A more accurate expression, covering a much larger range in $q$, was derived by \citet{eggleton83} and is based on an elegant fitting formula applied to the results of numerical integrations of the Roche-lobe volume:
\begin{equation}
f_{\rm E} = \frac{0.49 \, q^{2/3}}{0.6 \, q^{2/3}+\ln (1+q^{1/3})}.
\label{eqn:eggleton}
\end{equation}
For an extensive discussion of other formulations for the size of the Roche lobe see \citet{leahy}.

We start with the simpler, but more insightful, of the two expressions to derive the minimum period before Roche-lobe overflow commences.  Inserting the expression $a=R_{\rm L}/f_{\rm K}(q)$ from Eqns.~(\ref{eqn:rL}) and (\ref{eqn:kopal}) into Eqn.~(\ref{eqn:kepler}), we find
\begin{equation}
P = \frac{2 \pi}{G^{1/2}} \sqrt{\frac{81}{8}} R_{\rm L}^{3/2} M_{\rm com}^{-1/2} 
\label{eqn:Pmin_gen}
\end{equation}
which is independent of $M_{\rm host}$, and this is the motivation behind using the \citet{kopal59} formulation of $f_{\rm K}(q)$.  

The minimum orbital period will be attained when the orbit shrinks to the point where the companion radius equals $R_{\rm L}$, in which case we have
\begin{eqnarray}
P_{\rm min} & = &\frac{2 \pi}{G^{1/2}} \sqrt{\frac{81}{8}}R_{\rm com}^{3/2} M_{\rm com}^{-1/2} \nonumber \\
& \simeq &  8.85 \left(\frac{R_{\rm com}}{R_\odot}\right)^{3/2} \left(\frac{M_{\rm com}}{M_\odot}\right)^{-1/2} ~{\rm hr}
\label{eqn:Pmin_rm1}
\end{eqnarray}
When we do the calculations in this work, we will actually utilize the more accurate $f_{\rm E}(q)$ expression for the Roche lobe dependence.  However, we can still cast the expression for $P_{\rm min}$ explicitly as a function only of $R_{\rm comp}$ and $M_{\rm com}$, multiplied by a correction factor that is a very weakly dependent function of $q$: 
\begin{eqnarray}
P_{\rm min}  & \simeq & 8.85 \, \xi(q) \left(\frac{R_{\rm com}}{R_\odot}\right)^{3/2} \left(\frac{M_{\rm com}}{M_\odot}\right)^{-1/2} ~{\rm hr} \nonumber \\
{\rm with} ~~\xi(q) & \equiv & \left[\frac{f_{\rm K}(q)}{f_{\rm E}(q)} \right]^{3/2}
\label{eqn:Pmin_rm2}
\end{eqnarray}
A plot of the slowly varying function $\xi(q)$ is shown explicitly in Fig.~3 of \citet{nelson18}. 

Finally, the expression for $P_{\rm min}$, Eqn.~(\ref{eqn:Pmin_rm1}) can also be cast as a function of the density of the companion only.  The right hand side of that equation is manifestly in the form the inverse square root of the density of the companion star.  It can therefore be rewritten as
\begin{eqnarray}
P_{\rm min} & = & \sqrt{\frac{243 \pi}{8G}} \,  \bar{\rho}_{\rm com}^{~-1/2} \nonumber \\
& = &  10.45 \, \left(\frac{{\rm g~cm}^{-3}}{\bar{\rho}_{\rm com}}\right)^{1/2} ~{\rm hr}
\label{eqn:Pmin_rho1}
\end{eqnarray}
where $\bar{\rho}$ is the mean density.

\section{Mass-Radius Relation for Degenerate H-Rich Bodies}
\label{sec:rom}

\begin{figure}
\begin{center}
\includegraphics[width=1.02\columnwidth]{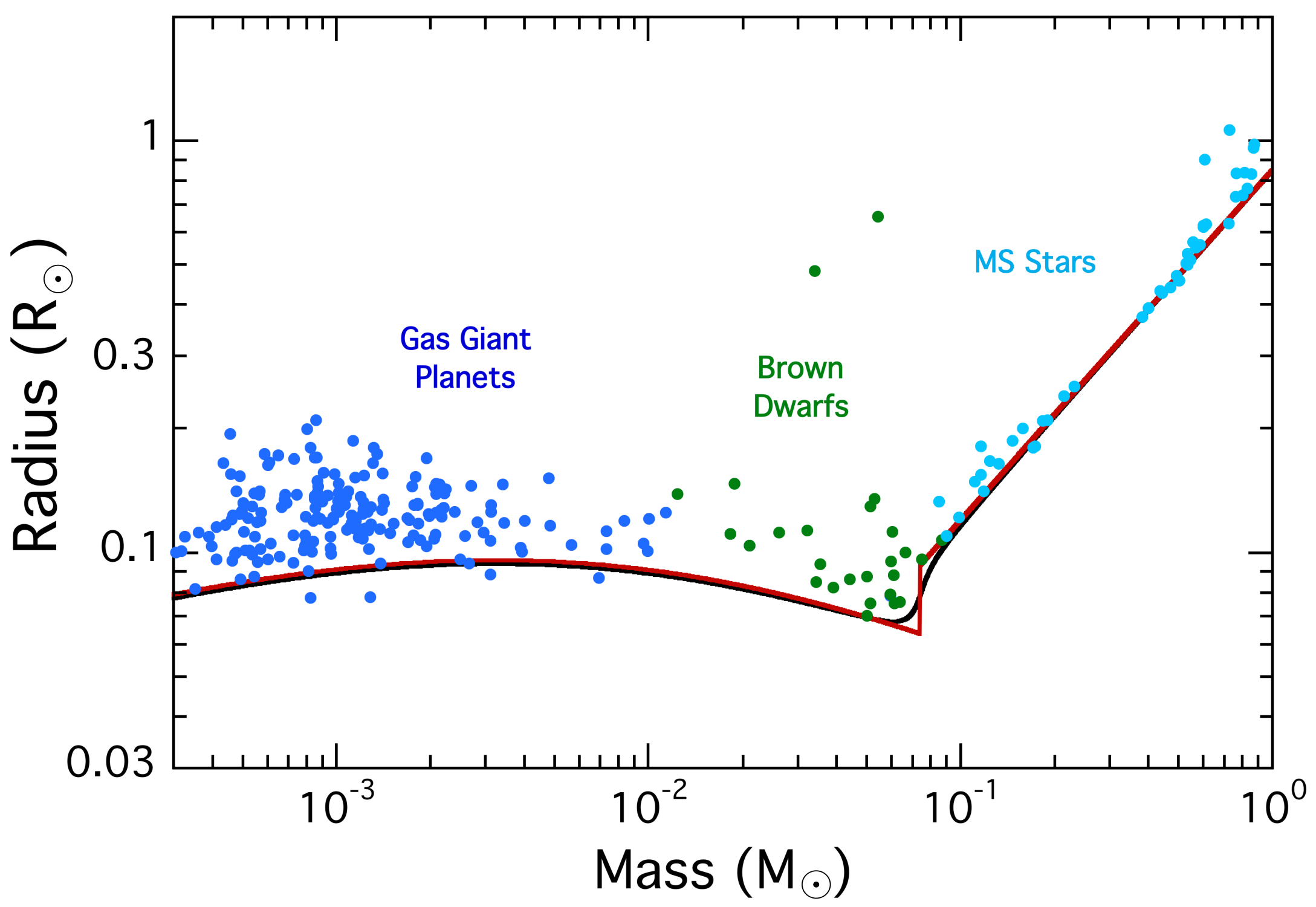}
\caption{Mass-radius relation for H-rich bodies spanning masses from Saturn to stars on the ZAMS up to 1 $M_\odot$ (red curve).  For the region between Saturn's mass and the end of the brown-dwarf range (at $\sim$0.074 $M_\odot$) we used equation (\ref{eqn:rofm}) with X = 0.7.   For stars on the ZAMS we use the simple expression given by equation (\ref{eqn:zams}).  The blue and cyan dots are an empirical sample of planets and main-sequence stars from \citet{chenkipping}.  The green points are our compilation of brown dwarfs taken from the literature (see Table \ref{tbl:BDs} and Sect.~\ref{sec:rom}). The black curve is an approximation to the red curve which has been smoothly blended near the transition between the ZAMS stars and brown dwarfs.  Note how both the red and black curves hug the lower locus of measured objects -- as desired.}
\label{fig:fig2} % Figure 2
\end{center}
\end{figure}

In order to estimate the radius of cold H-rich bodies we made use of the zero-temperature models of \citet{ZapolskySalpeter}.  These models represent the lower limit to the radius of brown dwarfs and planets for a specific mass and (homogeneous) composition.  We utilize an analytic expression devised by \citet{eggleton06} for the dependence of the radius on the mass and chemical composition of these objects:
\begin{equation}
R_{\rm com} \simeq 0.0128 (1+X)^{5/3} M_{\rm com}^{-1/3} \,g(M_{\rm com};X) 
\label{eqn:rofm}
\end{equation}
with
\begin{eqnarray}
g & = & g_1\cdot g_2 \\
g_1  &  = & \sqrt{(1-(M_{\rm com}/M_{\rm ch})^{4/3}} \\
g_2 & = &  \left(1+3.5(M_{\rm com}/M_{\rm p})^{-2/3}+M_{\rm p}/M_{\rm com}\right)^{-2/3} \\
M_{\rm ch} & = & 1.44 \,(1+X)^2 ~M_\odot\\
M_{\rm p} & = & 0.000566 \,(1+X)^{3/2} ~M_\odot
\end{eqnarray}
where $X$ is the H-mass fraction. In the expressions for $M_{\rm ch}$ and $M_{\rm p}$ we have simplified the original expressions of \citet{eggleton06} for the case of objects composed of H and He only.\footnote{The original expressions of \cite{eggleton06} were in terms of $Z_N$ and $A$, the atomic number and atomic weight, respectively, of each of the chemical constituents of the star.  For objects composed solely of H and He we derived an approximate weighting based on the value of $X$ only (see, \citealt{nelson03}).} We plot in Fig.~\ref{fig:fig2} the expression given by Eqn.~(\ref{eqn:rofm}) with $X = 0.7$ for masses between that of Saturn and the bottom of the zero-age main sequence ($M \simeq 0.074 \, M_\odot$). 

For masses above the bottom of the ZAMS, we use a simple $R(M)$ relation 
\begin{equation}
R_{\rm com}(M_{\rm com}) = 0.85 \, \left(M_{\rm com}/M_\odot\right)^{0.85} ~R_\odot
\label{eqn:zams}
\end{equation}
which is derived from a regression analysis applied to the lower main-sequence models of \citet{DNC} for the FGVH EOS \citep{FGVH} down to 0.085 M$_\odot$.  

Also plotted in Fig.~\ref{fig:fig2} are a sampling of planets, brown dwarfs, and lower main-sequence stars taken from a compilation of \citet{chenkipping}; and we augment this with our own compilation of brown dwarfs listed in Table \ref{tbl:BDs}. Because the distinction between gas-giant planets and brown dwarfs has been subject to considerable debate, we have included an extended discussion of this issue in the Appendix.

\begin{figure}
\begin{center}
\includegraphics[width=1.02\columnwidth]{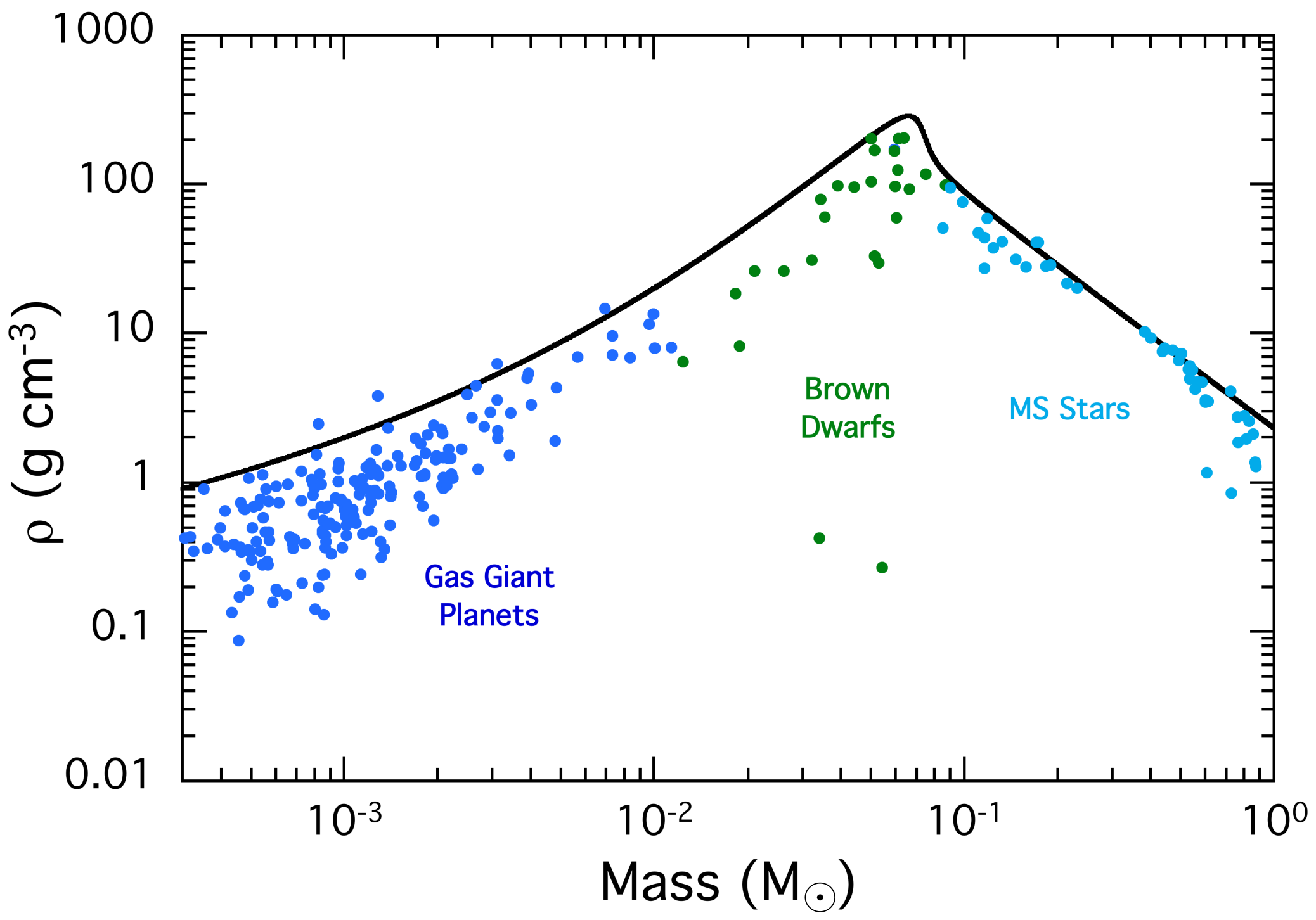}
\caption{Mean density-mass relation for H-rich bodies spanning masses from Saturn to stars on the ZAMS up to 1 $M_\odot$. The rest of the descriptors are the same as in Fig.~\ref{fig:fig2}. }
\label{fig:fig3} % Figure 3
\end{center}
\end{figure}

In Fig.~\ref{fig:fig3} we employ the same $R(M)$ relations used to construct the theoretical curve in Fig.~\ref{fig:fig2} to compute the mean density of the orbiting body as a function of its mass.  Here we also superpose the same collection of planets, brown dwarfs, and lower main-sequence stars compiled by \citet{chenkipping} and augmented with our list of BDs given in Table \ref{tbl:BDs}.  Note that, as desired, the curve forms a near upper boundary to the empirically observed systems. 

\begin{table}
\centering
\caption{Mass-Radius Pairs for Brown Dwarfs$^\dag$}
\begin{tabular}{lccc}
\hline
\hline
Name  & 	Mass  & 	Radius  & Reference  \\
            &   $M_J$  & $R_J$  &    \\
\hline 
SDSS J0857+0342 & $91.1\pm12.6$ & $1.07\pm0.04$ & 1 \\
Kepler-503      &  	$78.6 \pm 3.1$	&  $0.96^{+0.06}_{-0.04}$	 & 2	   \\	
WD1032+011 b  &	$69.6 \pm 6.4$ &  $1.0 \pm 0.1$ & 3 \\ 	
EPIC 201702477  & 	$66.9 \pm 1.7$	&  $0.757\pm0.065$	 &  4	   \\				
TOI-569         &   	$64.1\pm1.9$	&  $0.75\pm0.02$	 &    5    \\       	
WD 1202-024  b &      $63.9\pm10.5$  & $0.88\pm0.05$ & 6 \\									
CoRoT-15 b	&       $63.3^{+4.3}_{-4.1}$	&  $1.12^{+0.30}_{-0.15}$	 &  2   \\					
WASP-030        & 	$62.5\pm1.2$	&  $0.951^{+0.028}_{-0.024}$	 &    7    \\     		
KOI-415 b	        &       $62.1\pm2.69$	&  $0.79^{+0.12}_{-0.07}$	 &    2     \\							
V* V2384 Ori a	&       $56.7\pm4.8$	&  $6.52\pm0.33$	 &    8    \\  		
TOI-811 b	       &       $55.3\pm3.2$	&  $1.35\pm0.09$	 &     2    \\						
TOI-852 b	      &       $53.7\pm1.3$	&  $0.75\pm0.03$      &      2       \\							
TOI-503 b	      &       $53.6\pm1.1$	&  $1.29\pm0.30$	 &     9     \\  
SDSS J1411+2009 & $52.4\pm2.1$	& $0.70 \pm 0.04$ & 10 \\ 				
EPIC 212036875 b  & $52.3\pm1.9$	&  $0.874\pm0.017$	 &  11  \\ 			
TOI-1406 b	&       $46.0\pm2.7$	&  $0.86\pm0.03$	 &  2   \\					
Kepler-492 b	&       $40.8^{+1.1}_{-1.5}$	&  $0.82\pm0.02$	 &   12     \\  			
WASP-128 b	&       $37.19^{+0.83}_{-0.85}$	&  $0.94^{+0.22}_{-0.18}$	 & 2   \\					
EPIC 219388192 b &    $36.0\pm1.6$	&  $0.846\pm0.021$	 & 2   \\					
V* V2384 Ori b	&       $35.6\pm2.8$	&  $4.81\pm0.24$	 &    8  \\   		
NLTT 41135 b	&       $33.7\pm2.8$	&  $1.13^{+0.27}_{-0.17}$	 &     1    \\						
KELT-1 b	        &       $27.4 \pm 0.93$	&  $1.116^{+0.038}_{-0.029}$	 &     13   \\  		
CoRoT-3 b	&       $21.96\pm0.70$	&  $1.037\pm0.069$	 &  14  \\   	
GPX-1 b	        &       $19.7\pm1.6$	&  $1.47\pm0.16$	 &  2   \\					
Kepler-39 b	&       $19.1\pm1.0$	&  $1.11\pm0.03$	 &  12  \\ 				
HATS-70 b	&       $12.9^{+1.8}_{-1.6}$  &  $1.384^{+0.079}_{-0.074}$	 &  2       \\						
\hline\label{tbl:BDs}
\end{tabular}

{\bf Notes.} $^\dag$Brown dwarfs taken from the literature with masses measured to $\lesssim 15\%$ and radii with lower limits of $\lesssim 20\%$. (1) \citet{parsons12} (2) \citet{schneider11}; (3) \citet{}; (4) \citet{bayliss17}; (5) \citet{carmichael20}; (6) \citet{rappaport17b}; (7) \citet{triaud13}; (8) \citet{stassun06}; (9) \citet{subjak20}; (10) \citet{littlefair14}; (11) \citet{carmichael19}; (12) \citet{bonomo15}; (13) \citet{siverd12}; (14) \citet{southworth11}.
\end{table}

\section{Minimum Orbital Periods}
\label{sec:pmin}

\begin{figure}
\begin{center}
\includegraphics[width=1.02\columnwidth]{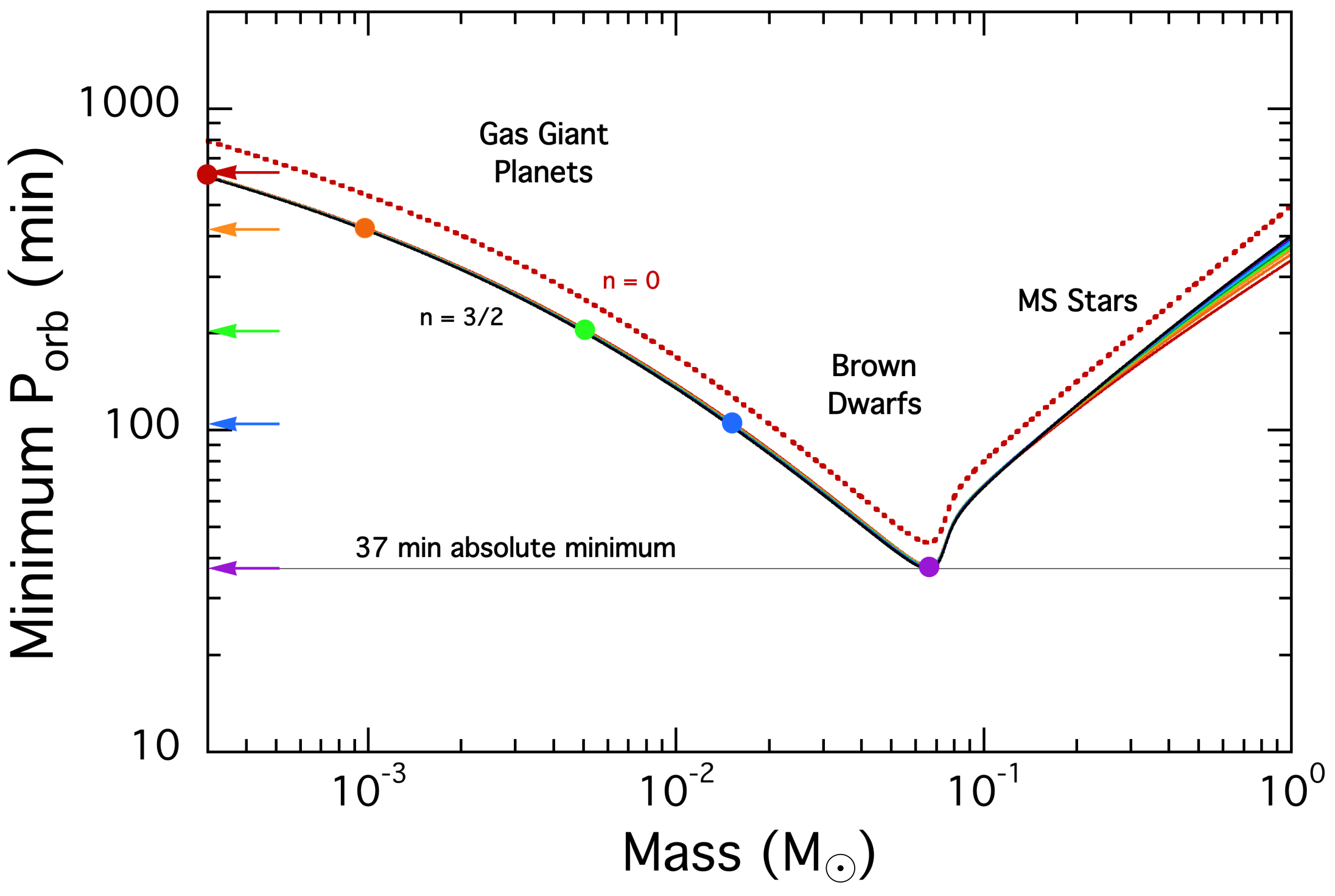}
\caption{Minimum allowed orbital period of H-rich bodies as a function of their mass. We have used the $R(M)$ relations displayed in Fig.~\ref{fig:fig2} and described by Eqns.~(\ref{eqn:rofm}) and (\ref{eqn:zams}), in conjunction with Eqn.~(\ref{eqn:Pmin_rm1}).  The various closely spaced colored curves (red, orange, ...\,blue, purple) are for different masses of the host star ranging from 0.3 $M_\odot$ to 1.4 $M_\odot$, respectively.  With the use of the approximate Eqn.~(\ref{eqn:Pmin_rm1}) all the curves would merge; but, not quite so with the more exact expression given by Eqn.~(\ref{eqn:Pmin_rm2}). The dotted red curve is the limit obtained for incompressible fluid bodies (see text). Heavy, filled, colored circles refer to fiducial-mass objects detailed in Table \ref{tbl:pofm}.}
\label{fig:fig4} % Figure 4
\end{center}
\end{figure}

We have used the $R(M)$ relations displayed in Fig.~\ref{fig:fig2} and described by Eqns.~(\ref{eqn:rofm}) and (\ref{eqn:zams}), in conjunction with Eqn.~(\ref{eqn:Pmin_rm2}) to derive the minimum allowed orbital period vs.~the body's mass.  The results are shown in Fig.~\ref{fig:fig4}.   

As we can see, there is a general trend of decreasing $P_{\rm min}$ from 620 min (10.3 hr) for Saturn-mass objects (red circle in Fig.~\ref{fig:fig4}), to 430 min (7.2 hr) for Jupiters (orange circle), to 104 min (1.7 hr) for objects on the boundary between super-Jupiters and brown dwarfs (blue circle), all the way down to 37 min (0.62 hr) for the coldest and most massive brown dwarfs (purple circle; see also \citealt{nelson18}).  These values are summarized in Table \ref{tbl:pofm}.  A simple fitting formula which is applicable for masses over the range $3 \times 10^{-4}-0.074 \, M_\odot$ is:
\begin{equation}
\ln P_{\rm min} \simeq 1.01-1.085 \ln m_{\rm com}-0.052 \ln^2m_{\rm com}
\label{eqn:Pmin_m}
\end{equation}
where $P_{\rm min}$ is in minutes, and $m_{\rm com} \equiv M_{\rm com}/M_\odot$.

\begin{figure}[b]
\begin{center}
\includegraphics[width=1.005\columnwidth]{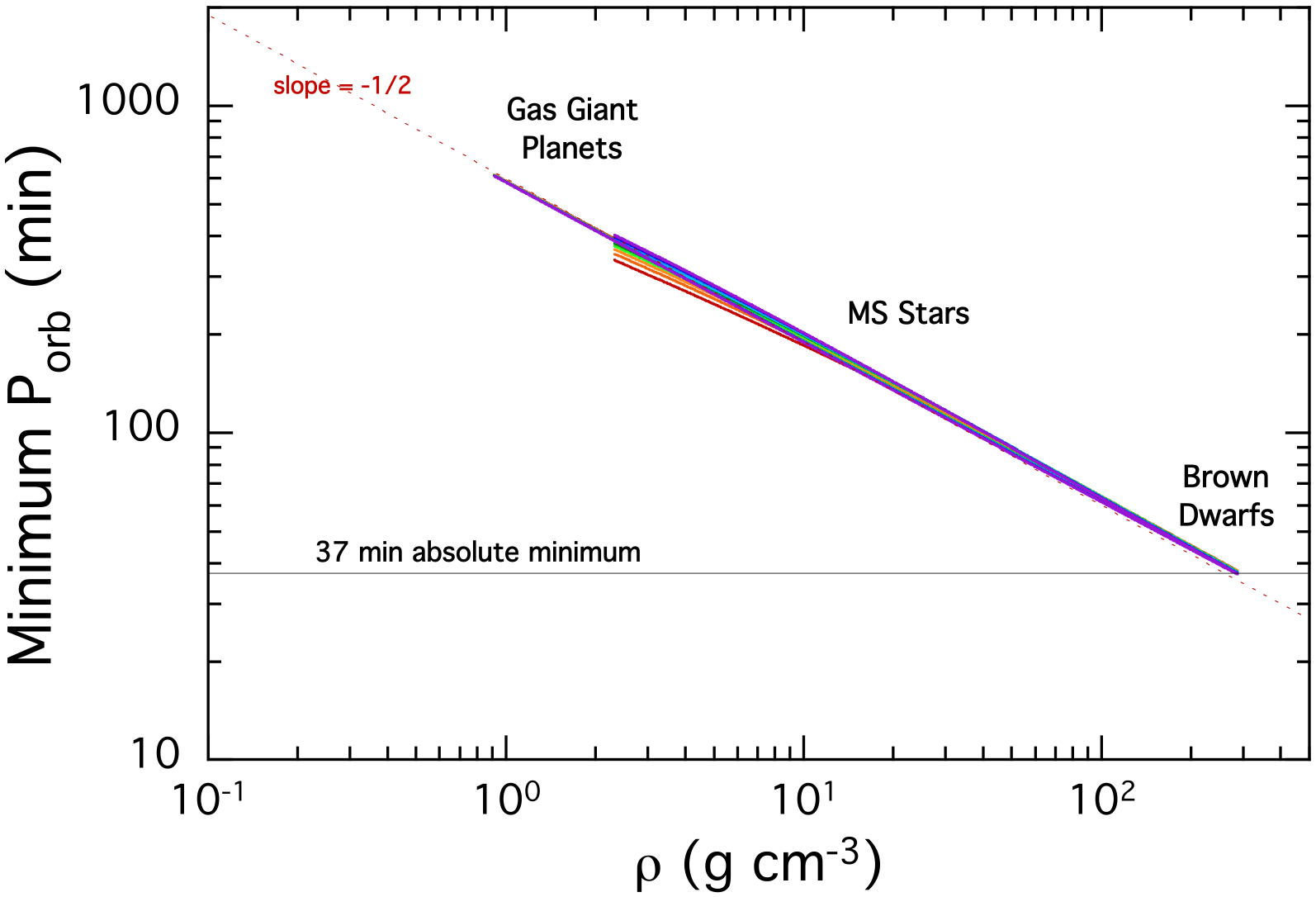}
\caption{Minimum allowed orbital period of H-rich bodies as a function of their mean density. Here we have used Eqn.~(\ref{eqn:Pmin_rho1}) which is derived from the approximate expression Eqn.~(\ref{eqn:Pmin_rm1}), but supplemented with the slowly varying function $\xi(q)$ defined in Eqn.~(\ref{eqn:Pmin_rm2}).  The various closely spaced colored curves  (red through purple) are for different masses of the host star ranging from 0.3 $M_\odot$ to 1.4 $M_\odot$, respectively.  With the use of the approximate Eqn.~(\ref{eqn:Pmin_rho1}) all the curves would merge; but, not quite so with the more exact expression given by Eqn.~(\ref{eqn:Pmin_rm2}). The appearance of the plot results from the fact that the curves start at the upper left (planets), decrease to the minimum period (brown dwarfs), and finally nearly retrace the same path back up toward the upper left (lower MS stars).}
\label{fig:fig5}
\end{center}
\end{figure} % Figure 5

Finally, we calculate $P_{\rm min}$ as a function of the mean density of the H-rich body.  For this, we use Eqn.~(\ref{eqn:Pmin_rho1}) multiplied by the function $\xi(q)$ given in Eqn.~(\ref{eqn:Pmin_rm2}).  The latter factor makes use of the more accurate Eggleton (1983) expression for the Roche-lobe radius.

The results for $P_{\rm min}(\bar{\rho}_{\rm com})$ are plotted in Fig.~\ref{fig:fig5}.  Here we see a nearly linear relation in the $\log(P_{\rm min})-\log \bar{\rho}_{\rm com}$ plane. The light dashed red line is a reference slope $-1/2$, as would be expected from Eqn.~(\ref{eqn:Pmin_rho1}).  The expression for that dashed line is given by:
\begin{eqnarray}
P_{\rm min}  \simeq  9.9  \, \left( \frac{{\rm g~cm}^{-3}}{\bar{\rho}_{\rm com}}\right)^{1/2} ~{\rm hr}
\label{eqn:Pmin_rho2}
\end{eqnarray}

We note that for main-sequence stars with mean densities between 2 and 10 g cm$^{-3}$ (corresponding to masses of 1.0 to 0.28 $M_\odot$) there is a small dispersion in $P_{\rm min}$ of $\pm 10\%$.  This is due to the fact that the \citet{kopal59} Roche lobe formula, which leads to the simple $\bar{\rho}_{\rm com}^{~-1/2}$ dependence, is not really applicable when the H-rich companion star is more massive than the host star (which we took to range from 0.3 to 1.4 $M_\odot$.  However, when we utilized the more exact Roche-lobe expression of \citet{eggleton83} we see the deviation from the simple $\bar{\rho}_{\rm com}^{~-1/2}$ dependence for the extremely low mass host stars.  Nonetheless, the overall $\bar{\rho}_{\rm com}^{~-1/2}$ dependence does an excellent job of representing $P_{\rm min}$ over two orders of magnitude in H-rich companion star mass and density.

\begin{table}
\centering
\caption{Minimum Orbital Periods of H-Rich Bodies}
\begin{tabular}{lcc}
\hline
\hline
 Object & $P_{\rm min}$ & $\langle \rho \rangle$  \\
  & (hr) & g cm$^{-3}$    \\
\hline
Saturn & 10.3 & 0.69 \\
Jupiter & 7.2 & 1.33 \\
5 $M_J$ & 3.3 & 8.69 \\
15 $M_J$ & 1.7 & 33.9 \\
Max brown dwarf & 0.62 & 280 \\
\hline
\label{tbl:pofm}
\end{tabular}

{\bf Notes.} Illustrative points taken from Fig.~\ref{fig:fig4}.
\end{table}

\section{Central Concentration of the H-Rich Body}
\label{sec:concentration}

\begin{figure}
\begin{center}
\includegraphics[width=\columnwidth]{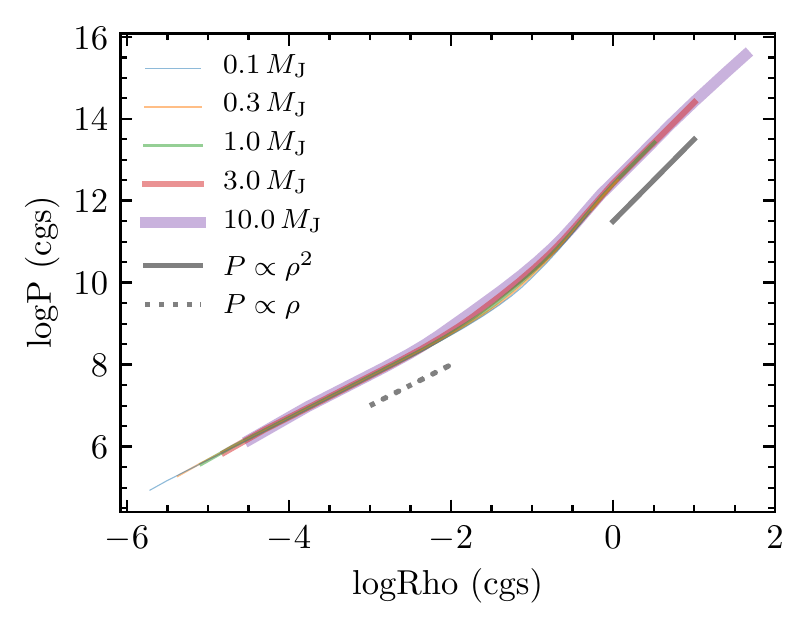}
\caption{Internal models for massive gas giant planets generated with {\tt MESA}. The models span the range from 0.1 $M_J$ to 10 $M_J$.  Most of the curves overlap to within the widths of the curves.  The solid and dashed line segments denote logarithmic slopes of +2 and +1, respectively.}
\label{fig:fig6} % Figure 6
\end{center}
\end{figure}

Thus far, we have been working under the assumption that the orbiting H-rich bodies are substantially centrally concentrated.  Bodies that might be roughly represented as $n = 3$ or even $n=3/2$ polytropes would qualify.  They have ratios of central to mean densities of 54 and 6, respectively.  For such centrally concentrated objects, we assume that the Roche-lobe formula, given by Eqn.~(\ref{eqn:eggleton}), is quite appropriate and accurate in terms of describing the size of the critical potential surface.

However, we know that massive gas giant planets and brown dwarfs are less centrally concentrated than an $n=3/2$ polytrope.   To make this more quantitative, we show in Fig.~\ref{fig:fig6} pressure-density curves, $P(\rho)$, for the interiors of cold H-rich bodies of masses of 0.1 $M_J$ to 10 $M_J$, in 5 logarithmic steps.  These models were generated with {\em MESA} \citep{MESAI,MESAII,MESAIII,MESAIV,MESAV} version r13573 using the included `make\_planets' test case.  The models assume a metallicity $Z = 0.02$, no solid core, and are shown at an age of 10 Gyr.  It is impressive that the curves nearly overlap, at least on this broad logarithmic scale.  The heavy black solid and dashed lines mark logarithmic slopes of 2 and 1, respectively.  And since most of the mass is represented fairly well by a $P \propto \rho^2$ relation, we conclude that these objects are more nearly represented by $n=1$ polytropes and less centrally concentrated than an $n=3/2$ polytrope.   

\begin{figure}
\begin{center}
\includegraphics[width=\columnwidth]{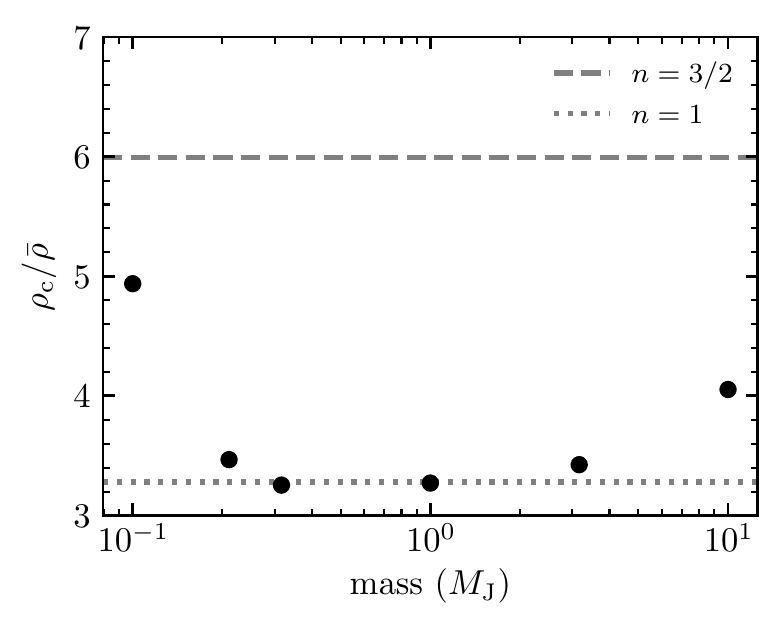}
\caption{Ratio of central ($\rho_c$) to mean ($\bar{\rho}$) density of the gas-giant models shown in Fig.~\ref{fig:fig6} plus one additional model at 0.2 $M_J$. Ratios of 3-4 are typical for these gas giants. }
\label{fig:fig7} % Figure 7
\end{center}
\end{figure}

In Fig.~\ref{fig:fig7} we show in more detail the ratio of central density to mean density for six different models over a wide mass range.  The geometric mean ratio of $\rho_c/\bar{\rho}$ is $\simeq 3.6$.  That ratio for an $n=1$ polytrope is 3.13.  Hence, we conclude the bulk of the non-main-sequence objects we are considering are well represented by $n=1$ polytropes.  

Unfortunately, to our knowledge, there are no equivalent expressions to Eqn.~(\ref{eqn:eggleton}) for $n=1$ polytropes with $\rho_c/\bar{\rho}$ is $\simeq 3.1$ that are filling their critical potential lobes.  This would be a good exercise for a self-consistent field calculation (see, e.g., \citealt{hachisu86}).  But, that is beyond the scope of this paper.  Thus, for now, we jump to a model that is of uniform density---and with a known solution.  This is the Roche limit for a uniform density, incompressible fluid body.  The Roche limit is usually expressed as $a_{\rm crit} = 2.44 R_{\rm host} ({\bar\rho}_{\rm host}/\bar{\rho}_p)^{1/3}$ where $a_{\rm crit}$ is the critical (i.e., minimum) orbital separation to avoid mass transfer, $R_{\rm host}$ is the radius of the host star, and the $\rho$'s are the mean densities of the host star and orbiting `planet' -- in this case the H-rich body.   

For our purposes in this paper, we can rewrite the Roche limit in the following form:
\begin{equation}
P_{\rm min} \simeq \sqrt{\frac{3 \pi (2.44)^3}{G \rho_{\rm com}}} = 12.6\, \left( \frac{{\rm g~cm}^{-3}}{\bar{\rho}_{\rm com}}\right)^{1/2} ~\,{\rm hr}.
\label{eqn:Pmin_rho3}
\vspace{0.1cm}
\end{equation}
This is directly analogous to Eqn.~(\ref{eqn:Pmin_rho1}), except that the leading coefficient here is somewhat larger.  This reflects the fact that the material in an incompressible configuration is less tightly bound than a compressible fluid of the same mean density.  

This relation for $P_{\rm min}$, using the results for an incompressible fluid ($n=0$), is shown as a dashed red curve in Fig.~\ref{fig:fig4}.  As we can see, it only raises the minimum allowed period by some 30\%.  We take this to be a firm upper limit on what the $P_{\rm min}(M_{\rm com})$ might be.  For most of the H-rich objects we are considering we surmise that the true answer lies somewhere between the two limiting curves (red and black in Fig.~\ref{fig:fig4}).

\section{Summary and Conclusions}
\label{sec:summary}

In this work we have examined how the minimum orbital period of cold H-rich bodies depends on the mass of the object, or alternatively, its density.  The basic conclusions are as follows.  (1) For any gas-giant planet or brown dwarf the minimum period is $\sim$37 minutes.  (2) For objects with orbital periods $\lesssim 100$ minutes we can conclude that we are observing a brown dwarf (or second WD) rather than a gas-giant planet.  More generally, (3) we give an approximate analytic expression (see Eqn.~\ref{eqn:Pmin_m}) for the minimum period as a function of the mass of the gas-giant planet or brown dwarf.  We can use this information to eliminate gas giant planet candidates with transit durations of 2-15 minutes and periods of $\lesssim 100$ minutes.  The same is true for brown dwarf candidates with periods $\lesssim 40$ minutes.

Our work makes use of the radius-mass relation, $R(M)$, for cold H-rich bodies based on the zero-temperature models of \citet{ZapolskySalpeter}. These models represent the lower limit to the radius of brown dwarfs and planets for a specific mass and (homogeneous) composition. For purposes of comparing this theoretical $R(M)$ relation with the empirical data, we have utllized the \citet{chenkipping} compilation of planets, brown dwarfs, and stars that has been augmented by our own compilation of a more complete list of 26 brown dwarfs.

\vspace{0.3cm}

\acknowledgments
%\section*{Acknowledgements}

J.S.~is supported by the A.F.~Morrison Fellowship in Lick Observatory and by the National Science Foundation through grant ACI-1663688.  L.N.~thanks the Natural Sciences and Engineering Research Council (NSERC Canada) for financial support through the Discovery Grants program.  This research has made use of NASA's Astrophysics Data System.

%L.N. thanks the Natural Sciences and Engineering Research Council (Canada) for financial support provided through a Discovery grant.  We also thank Calcul Qu\'ebec, the Canada Foundation for Innovation (CFI), NanoQu\'{e}bec, RMGA, and the Fonds de recherche du Qu\'{e}bec - Nature et technologies (FRQNT) for computational facilites.  Finally we thank J. Aiken for his technical assistance.



%\bibliographystyle{mnras}


\appendix
\vspace{4pt}
\centerline{\em Commentary on the Mass-Radius Relationship for Brown Dwarfs and Gas Giant Planets}

\vspace{4pt}
The issue of distinguishing brown dwarfs (BDs) from gas giant planets (GGPs) is very complex. Because of the importance of delineating these two populations, several attempts have been made to construct suitable definitions. In 2003, the Working Group on ExtraSolar Planets (WGESP) of the IAU proposed a working definition whereby the distinction was based solely on the mass of the object\footnote{IAU's working definition (immediately after 2003):\url{http://www.astro.iag.usp.br/~dinamica/WGEP.html}}. Specifically, the mass defining the boundary between GGPs and BDs was proposed as 13\,$M_J$ regardless of how the object was formed or its location \citep{boss07}. This definition has ostensibly arisen because a cloud of hydrogen-rich, solar-metallicity gas with a mass of 13\,$M_J$ (or higher) that undergoes gravitational collapse will fuse most of its primordial deuterium over an interval of $\lesssim 30$ Myrs (\citealt{nelson85}, 1986; \citealt{spiegel11}, and references therein). This phase is referred to as the deuterium-burning main sequence (DBMS). The upper end of the BD mass range is defined as the mass for which hydrogen-burning cannot achieve thermal quasi-equilibrium within a Hubble time. Higher mass objects that collapse as a result of the fragmentation of a hydrogen-rich gas cloud will be able to sustain nuclear fusion and thus become hydrogen-burning main sequence (HBMS) stars.

The formulation of precise definitions in astronomy (or any other natural science) can be a challenging task. For example, there is no absolutely precise definition of when a star is first on the zero-age main sequence (ZAMS). The reason for this is that stars never achieve complete thermal equilibrium (i.e., their gravo-thermal luminosity is never zero). This type of difficulty also arises in trying to establish the BD-MS star `boundary' because, although there is complete agreement that stars achieve approximate thermal equilibrium via sustained hydrogen burning, high-mass BDs can attain a substantial degree of thermal equilibrium within a Hubble time. Moreover, the treatment of the input physics (e.g., opacities, EOS) and the assumed metallicity can have a profound effect in establishing the boundary between BDs and MS stars when mass is taken to be the only determining criterion. For a solar metallicity, the uncertainty in the input physics places the upper limit on the BD mass to be between $\simeq 0.072$ to 0.080\,$M_\odot$.

The difficulty in establishing whether an object is a GGP or a BD is much more problematic. According to conventional wisdom, it is believed that GGPs form as a result of (cold) core accretion in a circumstellar dusty disk \citep{burgasser08}. Assuming that there is sufficient mass in the disk for accretion, it is thus possible for the mass of the object to grow much larger than 13 $M_J$ without undergoing deuterium burning (i.e., no DBMS phase). However, it is also possible that relatively massive objects can form in circumstellar disks by direct fragmentation of the disk. This process is very similar to the way in which stars and, by extension, BDs form\footnote{Such a pathway has been suggested for the giant planet GJ 3512B \citep{morales19}.}. Knowing the formation process would help in reaching a definitive conclusion, but given the absence of such knowledge, we must rely on observables such as the mass ($M$), radius ($R$), and $T_{\rm eff}$. Higher order observables that derive from multiband spectra and direct imaging of the atmosphere could provide the detailed atmospheric information (e.g., cloud structures, temperature profile, species differentiation, presence of grains, etc.) that is needed to make a more robust determination.

All of these issues have led to considerable debate on how GGPs should be distinguished from BDs or even if such a delineation should be made. \citet{hatzes15} (hereafter HR) make the claim that: ``objects with masses in the range $0.3\,M_J - 60\,M_J$ follow a tight linear relationship [in the $\log M - \log \rho$ plane] with no distinguishing feature to separate the low-mass end (giant planets) from the high-mass end (brown dwarfs)." They propose that all objects with masses in the range of $0.3 < M/M_J < 60$ should be viewed as GGPs. This definition is based on the linearity (and continuity) of the slope in the mass-density relationship (see their Figure 1). This is a purely phenomenological definition and thus does not address the underlying physics of these objects. Moreover, the paucity of data in the range of $20 \, M_J - 40 \,M_J$ makes this type of analysis particularly challenging. Our Figure 2 (see also Table 1) contains significantly more data than was available to HR and reveals that the assumption of a linear mass-radius (MR) relationship (in the logs) should be re-examined. Our red curve (zero temperature hydrogen-rich models) shows that the slope of the MR relationship changes considerably between masses of 0.3 $M_J$  to $\simeq 0.072\,M_\odot$ (i.e., just below the hydrogen-burning minimum mass [HBMM]. Near the HBMM, we would naively expect the $M-R$ exponent ($R \sim M^\xi$) to be approximately -1/3 because the EOS is dominated by nonrelativistic electron degeneracy. However, even at this upper end of the mass range for BDs, the effects of Coulombic interactions start to become significant and this has the effect of making $\xi$ more positive (i.e., flattening the $M-R$ relationship). For even lower masses, other effects such as Thomas-Fermi corrections that more accurately account for electron-nucleus interactions, in addition to exchange effects and correlation energies (see, e.g., Seager et al. 2008) further increase the value of $\xi$.

For chemically homogeneous, zero-temperature objects, the Zapolsky-Salpeter (1969) models show that $\xi$ changes from approximately -1/3 to +1/3 as the mass decreases to terrestrial values. In examining the observational data in Figure 2, we see that the smallest GGPs and BDs (in radius) tend to follow the theoretical predictions reasonably closely (red curve). There is considerable scatter in the BD mass range but this is to be expected because these BDs are discovered at various stages in their contraction (BDs have very long Kelvin-Helmholtz times and thus their radii tend to be reasonably sensitive to their ages). Nonetheless, there is a reasonably pronounced dip in the radii (compared to the radii of MS stars) at $M \simeq 0.07 M_\odot$. This feature is completely consistent with the HBMM inferred from theoretical models of Population I objects.  Moreover, the lower envelope of $R-M$ observations is not inconsistent with the zero-temperature theoretical models that predict  the lower limit for the radii of objects of a specific mass (and chemical composition). We emphasize that in the absence of any other information concerning the correlation between radius and mass, it is reasonable to carry out a linear regression (i.e., to assume a constant $\xi$); however, the changing `physics' of these objects (with mass) requires that $\xi$ be treated as variable. Thus we believe that the claim of Hatzes \& Rauer that: (1) BDs should be subsumed into ``the upper end of the giant planet sequence''; and, (2) the boundary between GGPs and MS stars should be set equal to 0.06\,$M_\odot$ appears to be an oversimplification. 

\citet{chenkipping} have re-examined/refined the work of HR with respect to the mass limits of `Jovian worlds'. Based on their analysis, this limit extends from $0.41\,M_J$ to 80\,$M_J$ (higher-mass objects are defined to be stars). They claim that: ``There is no discernible change in the $M-R$ relation from Jupiter to brown dwarfs. Brown dwarfs are merely high-mass planets, when classified using their size and mass''. This conclusion is largely based on a linear regression analysis carried out on GGP and BD data (see their Figure 3). They find a relatively flat dependence of radius on mass ($R \sim M^{0.04}$). Again, the challenge with this type of purely empirical analysis is that brown dwarfs are rare; the \citet{chenkipping} analysis includes 150 objects in the mass range corresponding to giant planets, and only 5 brown dwarfs (two of which have such high masses that their status as brown dwarfs becomes questionable). Therefore any regression over the entire range of mass will be dominated by the properties of the giant planets. Moreover, because the mass limits of 0.41\,$M_J$ to 13\,$M_J$ straddle the mass corresponding to the extremum in the radius of the zero-temperature models (that mass is approximately the geometric mean of the limits), we would naturally expect any regression to show a relatively flat $M-R$ relationship. The fact that a power law can provide an acceptable fit to the $M-R$ relationship over this mass range does not imply that objects in this mass range are indistinguishable.

The bottom line is that the theoretical models (for a given age and ignoring the DBMS) imply that there should be no abrupt change in the $M-R$ relation exponent ($\xi$) across any reasonable boundary that is chosen to delineate GPPs from BDs (with the proviso that the chemical structure of the objects is similar). This continuity in $\xi$ does not, however, imply that BDs should subsumed into the class of GGPs nor does it imply that an extension of a linear fit to the MR relationship determined for GGPs can be reasonably applied to BDs. For example, the $M-R$ data shown in Figure 2 are not inconsistent with the simple  \citet{ZapolskySalpeter} models (red curve), and those models show that $\xi$ can change significantly over the nominal BD mass range. Moreover, if objects in this mass range are not formed by core accretion and undergo a DBMS phase (even if it only lasts for $< $ 100 Myr), they are clearly a separate class from GGPs because they formed similarly to stars and achieved approximate thermal equilibrium via nuclear fusion.

%%%%%%%%%%%%%%%%%%%%%%%%%%%%%%%%%%%%%%%%%%%%%%%%%%

%%%%%%%%%%%%%%%%%%%%%%%%%%%%%%%%%%%%%%%

\end{document}